\documentclass{article}    % Specifies the document style.

\title{An Accelerating Universe Around A Blackhole}  % Declares the document's title.
\author{A. M. Harunar Rashid\footnote{ Department of Physics, Dhaka University,Dhaka, Bangladesh},  A. Momen\footnote{Department of Physics, Dhaka University,Dhaka, Bangladesh} \footnote{email:arshadmomen@yahoo.com}, and A. L. Choudhury \footnote{ Dept. of Chemistry and Physics, Elizabeth City State University, Elizabeth City, NC 27909, USA} \footnote{ Fulbright Scholar, Department of Physics, Dhaka University, Dhaka , Bangladesh.}\footnote{email: alchoudhury@mail.ecsu.edu}
}         % Declares the author's name.

\begin{document}           % End of preamble and beginning of text.

\maketitle                 % Produces the title.
\begin{abstract}
{We assume the creation of a blackhole in a physical universe. We now conjecture that this blackhole will then separate itself from the physical universe and build up an extra dimensional entity associated with the physical universe. We postulate this extra dimensional entity is to be orthogonal to the physical universe. We further conjecture that this blackhole is a Schwartzschild blackhole. We assume also that this physical universe and the blackhole together span a seven dimensional space with a common time coordinate. We then generate the Einstein equation. Using the time-blackhole-radial and the time-time component of the equation we show that the Hubble parameter is positive and time dependent if we conjecture that both the scale factor and the radius of the blackhole reduce exponentially. Under the same assumption we also calculate the deacceleration parameter and show that under certain constraint on the parameters the universe accelerates.}
\end{abstract}
\section{Introduction} 
{ \quad Recently Choudhury [1] has shown that the acceleration of the physical universe can originate from an adiabatic pressure generated from a contracting blackhole. The experimental signal [2] that the universe is accelerating has been observed from supernova data.This signal forced the cosmologists to reintroduce the cosmological constant in the Einstein equation to account for the generation of the acceleration. The introduction of this constant leads to the postulation of the existence of dark energy in cosmology. However the cosmological constant appears to us not sufficient as a dynamical process to explain such a significant phenomenon as acceleration of the universe. Choudhury and Pendharkar [3] first postulated that the pressure generated from a wormhole of the type introduced by Gidding and Strominger [4] [5] can cause such an accelerated expansion of the universe which surrounds the wormhole.}
{\par  Since the hypothetical wormhole is not a real entity, Choudhury [1] looked for other causes that might explain the  dynamical generation of acceleration. During the process of the formation of a blackhole, Choudhury postulated that the contracting gas develops pressure in accordance with adiabetic thermodynamical gas laws. This pressure then influences the surrounding physical universe. This semiclassical model can then generate acceleration in the expanding universe. The success of this model led to a search for the improved treatment of this concept.
\par Earlier Choudhury[6] had developed a model of higher dimension in which he assumed the time-space is seven dimensional[7]. He has introduced in the extra three dimension a Gidding-Strominger-Maeda wormhole, which is by construction expanding. He then generated a classical pressure in the wormhole assuming the adiabetic gas law to be valid in the wormhole. This pressure influences the physical universe contained in the other three dimensions. As a result the physical universe expands and accelerates at a fluctuating rate.
\par We wish to use a similar concept in a new model. Instead of putting a wormhole in the extra dimensional space, we assume a blackhole there. A blackhole generated in a physical universe creates its own event horizon keeping everything hidden within the horizon except through a very negligible Hawking radiation. Hence the blackhole can safely be conjectured to be within the extra dimensional space whose coordinates are orthogonal to the original physical world. With such a conjecture we assume that the physical universe is a Friedman-Robertson-Walker space and the detached arena is occupied by a Schwarzschild blackhole occupying the extra dimensional space. We also introduce two scale factors $ \alpha(t)$ for the physical universe and $\beta(t)$ for the Schwarzschild space.
\par The separation of the universe into a Schwarzschild space and a Friedman-Walker-Robertson physical space is not in conflict with the perfect cosmological principle which states that the smoothed out universe is unchanging as well as spatially homogeneous. We are postulating that, because of the singular character of the blackhole, the space surrounding it has a different coordinate grid than that of the physical world and this simple assumption leads from Einstein equations to an accelerating universe which is experimentally observed.  
\par We then set up the Ricci tensors and obtain the relations in section 2, which result from the Einstein equation. We have also set up the energy momentum tensors. The eight different equations we obtain are not very easy to solve simultaneously. In section 3 we have set up some physical criteria for the scale factor $\beta(t)$ and blackhole radius $r_\beta$ in the Schwarzschild sector. We assume there that both $\beta(t)$ and $r_\beta$ shrink exponentially but at a different rate. Here $\beta(t)$ shrinks faster than the radius $r_\beta$.
\par Equipped with this assumption  and setting for simplicity $k_\alpha = 0$, the Hubble parameter can be calculated for large values of $t_0$. We then calculate the time dependent deacceleration parameter. Under a certain constraint we show that the deacceleration parameter becomes negative. This signature tells us that the universe is accelerating.}

\section{Einstein Equation}
{\quad We are assuming that the physical universe along with a blackhole behaves like a space with the physical universe spanning a three dimensional space and the blackhole the other three dimension with a total space dimension of six. Bringing in a common time for both, our space-time dimension becomes seven. The physical universe part should be a Friedman-Robertson-Walker space, whereas the other 1 + 3 space-time contains the Schwarzschild blackhole. We now conjecture that in this super-imposed space-time, the interval is given by the relation}
\begin{equation}
d\tau^2 = d{\tau_p}^2+d{\tau_b}^2
\end{equation}
{where}
\begin{equation}
d{\tau_p}^2 =-dt^2+{{\alpha}}^2(t)(\frac{dr_\alpha^2} {1-{k_\alpha} {r_\alpha}^2}+{r_\alpha^2}d{\Omega_\alpha}^2),
\end{equation}
{and}
\begin{equation}
d{\tau_b}^2 =-\beta(t)^2[(1-{\frac{2m} {{r_\beta}}})dt^2+(1-{\frac{2m}{{r_\beta}^2}})^{-1}dr_\beta^2+{r_\beta^2}d{\Omega_\beta}^2]. 
\end{equation}
{In the above equations $\alpha(t)$ and $\beta(t)$ are two scale factors. We assume c=1 and we write}
\begin{equation}
d\tau^2=g_{\mu\nu}d{x^\mu}d{x^\nu},
\end{equation}
{with}
\begin{equation}
x^0=t, x^1=r_\alpha, x^2= \theta_\alpha, x^3=\phi_\alpha, x^4=r_\beta, x^5=\theta_\beta, and\quad x_6=\phi_\beta.
\end{equation}
{where the suffix $\alpha$ refers to the physical world and $\beta$ to the blackhole. We have therfore,}
\begin{equation}
g_{tt}= -[1+\beta^2(t)(1-\frac{2m} {r_\beta})], 
\end{equation}
\begin{equation}
g_{{r_\alpha}{r_\alpha}}=\frac{\alpha^2(t)} {1-{k_\alpha}{r_\alpha^2}},
\end{equation}
\begin{equation}
g_{{\theta_\alpha}{\theta_\alpha}}= \alpha^2(t) {r_\alpha}^2
\end{equation}
\begin{equation}
g_{{\phi_\alpha}{\phi_\alpha}}= \alpha^2(t) {r_\alpha}^2 sin^2{\theta_\alpha} ,
\end{equation}
\begin{equation}
g_{{r_\beta}{r_\beta}}= \beta^2(t)(1-\frac{2m} {r_\beta})^{-1} ,
\end{equation}
\begin{equation}
g_{{\theta_\beta}{\theta_\beta}}= \beta^2(t) {r_\beta}^2 ,
\end{equation}
{and}
\begin{equation}
g_{{\phi_\beta}{\phi_\beta}}= \beta^2(t) {r_\beta}^2 sin^2{\theta_\beta} .
\end{equation}
{ All non-diagonal components of g are zero. We can easily verify that}
\begin{equation}
g^{ii}=\frac 1 g_{ii}.
\end{equation}
{ Using the definition of Riemann's curvature tensor}
\begin{equation}
R^{\alpha}_{\beta\gamma\delta}= \frac{{\partial\Gamma^\alpha}_{\beta\delta}} {\partial x^\gamma}-\frac{{\partial\Gamma^\alpha}_{\beta\gamma}} {\partial x^\delta}+{ \Gamma^\alpha}_{\lambda\gamma}{\Gamma^\lambda}_{\beta\delta}-{\Gamma^\alpha}_{\lambda\delta}{\Gamma^\lambda}_{\beta\gamma}
\end{equation}
{with the connection}
\begin{equation}
\Gamma^{\alpha}_{\beta\gamma}={\frac {1}{ 2}} {g^{\alpha\lambda}}(\frac{\partial g_{\lambda\beta}} {\partial x^\gamma}+\frac{\partial g_{\lambda\gamma}} {\partial x^\beta}-\frac{\partial g_{\beta\gamma}} {\partial x^\lambda}),
\end{equation}
{we can now calculate the Ricci tensors. The nonvanshing ones turn out to be}
\begin{equation}
R_{tt}=-\frac{3\ddot\alpha} \alpha +\frac{2\dot\alpha\dot\beta(r_\beta-2m)}{\alpha{}f(r_\beta,\beta)}-\frac{3\ddot\beta} \beta+\frac{3{\dot\beta}^2(r_\beta-2m)} {f(r_\beta,\beta)}-\frac{\beta^2 m^2 (r_\beta-2m)} {r_\beta^4 f(r_\beta,\beta)}+A(r_\beta,m),
\end{equation}
{where}
\begin{equation}
A(r_\beta,m)=\frac{m(r_\beta-2m)} {r_\beta^4}-\frac{m^2}{r_\beta^2(r_\beta-2m)}.
\end{equation}
{We have also set}
\begin{equation}
f(r_\beta, \beta)=r_\beta +{\beta}^2(r_\beta-2m).
\end{equation}
{Similarly}
\begin{equation}
R_{{r_\alpha}{r_\alpha}}={\frac{r_\beta} {(1-{k_\alpha}{r_\alpha}^2)}}+D(\alpha,\beta),
\end{equation}
{where}
\begin{equation}
D(\alpha,\beta)=\frac{\ddot\alpha \alpha} {f(r_\beta,\beta)}+\frac{\dot\alpha \alpha \dot \beta \beta[1-2(r_\beta-2m]} {f^2(r_\beta,\beta)}+ \frac{3\dot\beta{\dot\alpha}\alpha}{\beta{f(r_\beta,\beta)}} +\frac{{\dot\alpha}^2} {f(r_\beta,\beta)}+\frac{k_\alpha}{r_\beta},
\end{equation}
\begin{equation}
R_{{\theta_\alpha}{\theta_\alpha}}={r_\alpha}^2{r_\beta}[{\frac{({\ddot\alpha}f(r_\beta,\beta)-{\dot\alpha}\alpha{\dot\beta}\beta(r_\beta-2m))}{f^2(r_\beta,\beta)}}+\frac{2{\dot\alpha}} {f(r_\beta,\beta)} +\frac{3\alpha{\dot\alpha}{\dot\beta}} {\beta f(r_\beta,\beta)}],
\end{equation}
\begin{equation}
R_{{\phi_\alpha}{\phi_\alpha}}={r_\beta}{{r_\alpha}^2}{sin^2{\theta_\alpha}}[\frac{\alpha{\ddot\alpha}} {f(r_\beta,\beta)}+\frac{2\dot\beta\alpha\dot\alpha} {{\beta}f(r_\beta,\beta)}+\frac{\alpha \dot\alpha \dot\beta r_\beta}{f(r_\beta,\beta)}-\frac{2(1-3k_\alpha {r_\alpha}^2)} {{r_\alpha}^2 r_\beta}]
\end{equation}
\begin{equation}
R_{{r_\beta}{r_\beta}}= H_1 + H_2 + H_3
\end{equation}
{where}
\begin{equation}
 H_1 =\frac{(\ddot\beta+{\dot\beta}^2){r_\beta}^2 f(r_\beta,\beta)- 2{(\beta\dot\beta)^2}{r_\beta}^2 (r_\beta-2m)+{\dot\beta}^2 {r_\beta}^2 f(r_\beta,\beta)} {{f^2}(r_\beta,\beta)(r_\beta-2m)},
\end{equation}
\begin{equation}
H_2 = \frac{m\beta^2[2r_\beta+2\beta^2(r_\beta-m)]+\beta\dot\beta {r_\beta}^4-m^2\beta^4} {{r_\beta}^2f^2(r_\beta,\beta)},
\end{equation}
{and}
\begin{equation}
H_3=\frac{3{\dot\alpha}\beta{\dot\beta}{r_\beta}^3-m{\beta}^2 \alpha} {\alpha{r_\beta} f(r_\beta,\beta)(r_\beta-2m)}+\frac{ m} {r_\beta(r_\beta-2m)^2}.
\end{equation}
\begin{equation}
R_{{\theta_\beta}{\theta_\beta}}= {r_\beta}^3 \frac{{\ddot\beta} f(r_\beta,\beta)+{\dot\beta}^2 r_\beta}{f^2(r_\beta,\beta}+\frac{m\alpha+3{r_\beta}^3 \dot\alpha \beta \dot\beta} {\alpha f(r_\beta,\beta)}+\frac{2m(r_\beta-3m)} {r_\beta(r_\beta-2m)}.
\end{equation}
\begin{equation}
R_{{\phi_\beta}{\phi_\beta}}= {r_\beta}^3 {sin^2}\theta_\beta[\frac{\beta \ddot\beta+2{\dot\beta}^2} {f(r_\beta,\beta)}+\frac{\dot\alpha \beta \dot\beta} {\alpha}-\frac{{\dot\beta}^2 {r_\beta}}{f^2(r_\beta,\beta)}+\frac{m} {f(r_\beta,\beta) {r_\beta}^3} +\frac{2m^2} {{r_\beta}^4 (r_\beta-2m)}].
\end{equation}
{The only non-diagonal component is $R_{tr_\beta}$ which is given by}
\begin{equation}
R_{tr_\beta}=\frac{3\dot\alpha m{\beta}^2} {\alpha r_\beta f(r_\beta,\beta)}+\frac{2m{\dot\beta }\beta} {{r_\beta} f(r_\beta,\beta)}+\frac{\dot\beta} {\beta r_\beta}
\end{equation}
{The Einstein equation is}
\begin{equation}
R_{\mu\nu}=-{\frac{8\pi} 3}S_{\mu\nu},
\end{equation}
{where}
\begin{equation}
S_{\mu\nu}=T_{\mu\nu}-{\frac2 7} g_{\mu\nu}T,
\end{equation}
{The quantity $T_{\mu\nu}$ is given by the relation}
\begin{equation}
T_{\mu\nu}=Pg_{\mu\nu}+(P+\tilde{\rho})U_\mu U_\nu.
\end{equation}
{In the above expression}
\begin{equation}
U_0=\surd(1+{\beta}^2(1-\frac{2m} {r_\beta});\quad U_1=U_2=...=U_6;
\end{equation}
{and}
\begin{equation}
P=Gp_\alpha+G'p_\beta,\quad and \quad \tilde{\rho}=G\rho_\alpha+G'\rho_\beta.
\end{equation}
{The quantities $p_\alpha$ and $\rho_\alpha$ are the pressure and density in the physical world and quantities $p_\beta$ and $\rho_\beta$ are the corresponding expressions for the blackhole region. Similarly G and G' are  the gravitational constants for those regions.
\par The value of T yields}
\begin{equation}
T=-\tilde{\rho}+6P.
\end{equation}
{Hence}
\begin{equation}
S_{tt}={\frac{f(r_\beta,\beta)}{7r_\beta}}[9\tilde{\rho}+12P]=Q,
\end{equation}
{and}
\begin{equation}
S_{ii}=Pg_{ii},
\end{equation}
{where i stands for all spacial co-ordinates.}
\section{Asymptotic Solution Of The Einstein Equation}
{\quad\quad The Einstein equations are very involved and to obtain a satisfactory solution we have to carry out numerical calculation which will be very complicated. However, in order to get an idea about the combined behavior we want to carry out an approximate solution with some physical underpinning on the blackhole in two steps. We notice that the blackhole involves two quantities, the scale function $\beta(t)$ and $r_\beta$, the radius of the blackhole. Here we introduce a physical constraint that reduces exponentially. Similarly we assume that the radius $r_\beta$ shrinks gradually. However we assume their exponential reduction are happening at a different rate.
\par More rigorously, we assume that the time dependence of $\beta(t)$ satisfies the relation:}
\begin{equation}
\beta(t)={\beta_0}e^{-{\Delta}t},
\end{equation}
{where both $\beta_0$ and $\Delta$ are two positive parameters. Similarly we assume that the radius $r_\beta$ does also shrink exponentially according to the following relation}
\begin{equation}
r_\beta=r_0 e^{-{\Delta}'t}
\end{equation}
{\par For future use we conjectur that the quantities $\Delta$ and ${\Delta}'$ would have to satisfy the following relation:}
\begin{equation}
\Delta>{\Delta}'.
\end{equation}
{We interpret the above relation by saying that the size of the blackhole dynamically follows the scale function but at a different rate.}
\section{Hubble Parameter}
{\quad  From the definition of the time dependent Hubble parameter for the physical universe given by the relation:}
\begin{equation}
H_\alpha (t)= \frac{{\dot\alpha}(t)} {\alpha(t)},
\end{equation}
{and from the Einstein equation}
\begin{equation}
R_{tr_\beta}=0,
\end{equation}
{we get from Eq.(29) for finite time}
\begin{equation}
{H_\alpha}(t)=-\frac{[2m{\beta}^2+r_\beta+{\beta}^2(r_\beta-2m)]{\dot\beta}} {3m{\beta}^3}.
\end{equation}
{Using Eqs.(38)and (39) for very large $t=t_0$, we get}
\begin{equation}
H_\alpha(t_0)={\frac{r_0 \Delta} {3m{\beta_0}^2}}[e^{(2\Delta-\Delta')t_0} +{\beta_0}^2 e^{-\Delta't_0}].
\end{equation}
{\quad This implies that the quantity $H_\alpha(t_0)$ is always positive. Hense the physical universe expands and is time depended.} 
\section{Deacceleration Parameter}
{\quad From Eq.(30) we find}
\begin{equation}
R_{tt}=-{\frac{8\pi} 3}S_{tt}
\end{equation}
{Using Eqs. (16) and (36) we find}
\begin{equation}
-\frac{3\ddot\alpha} \alpha +\frac{2\dot\alpha\dot\beta(r_\beta-2m)}{\alpha{}f(r_\beta,\beta)}-\frac{3\ddot\beta} \beta+\frac{3{\dot\beta}^2(r_\beta-2m)} {f(r_\beta,\beta)}-\frac{\beta^2 m^2 (r_\beta-2m)} {r_\beta^4 f(r_\beta,\beta)}+A(r_\beta,m)=-\frac{8\pi Q}{3}.
\end{equation}
{For small $p_\alpha$ and $p_\beta$, we assume}
\begin{equation}
P \rightarrow 0.
\end{equation}
{For fixed $r_\beta$ and $\beta \rightarrow 0$, we get}
\begin{equation}
\frac{f(r_\beta,\beta)}{r_\beta}=\frac{r_\beta+{\beta}^2(r_\beta-2m)} {r_\beta} \rightarrow 1.
\end{equation}
{Therefore}
\begin{equation}
Q \rightarrow {\frac9 7} {\tilde \rho}={\frac9 7}[G\rho_\alpha+G'\rho_\beta] \rightarrow {\frac9 7}G'\rho_\beta,
\end{equation}
{because the density of the blackhole is the dominating factor.We can convert $\rho_\beta$ into the form}
\begin{equation}
\rho_\beta=\frac{3m} {4\pi\beta^3(t){r_\beta}^3}.
\end{equation}
 {From the definition of the deacceleration parameter of the physical universe, we know for large $t_0$}
\begin{equation}
q_0(t_0)=-{\frac{{\ddot\alpha}(t_0)} {\alpha(t_0)}}{\frac{1} {{H_\alpha}^2(t_0)}}.
\end{equation}
{Incorporating the values of $\beta(t)$ and $r_\beta(t)$ from Eqs.(38) and (39), $\rho_\beta$ from Eq.(14)and suppressing the terms which becomes negligible for large $t_0$, we find}
\begin{equation}
q_0(t_0)={\frac{3m^2{\beta_0}^4} {{\Delta}^2{r_0}^2}}{e^{-(\Delta-5\Delta')t_0}}[-{\frac{2m^3} {r_0^5}}e^{(-5\Delta+2\Delta')t_0}-{\frac{2m^2} {r_0^4}}e^{(-3\Delta+\Delta')t_0}-\frac{18G'm}{7\beta_0^3 r_0^3}]
\end{equation}
{Under the condition $\Delta >\Delta'$, we notice that $5\Delta>2\Delta'$ and $3\Delta>\Delta'$. Therefore for large $t_0$ the last term within the bracket in the above equation dominates and $q_0$ becomes negative. As a consequence the the universe accelerates.}
\section{Concluding Remarks}
{\quad We assumed that a blackhole in an advanced stage detaches itself from the physical universe and  enters into an associated extra dimensions. We postulated that the physical universe retains its Friedman-Robertson-Walker nature, whereas, the blackhole in the extra dimension takes a Schwarzschild character. Together they generate a seven dimensional space. In this space we set up the Einstein equation. The eight involved equations are quite complicated. Instead of looking for a rigorous solution we decided to solve the  pertinent equations which yield the Hubble and the deacceleration parameter under some physical conjecture. We assumed that both the scale factor of the blackhole space and its radius reduce at an exponential rate. The rate of reduction of the scale factor is assumed to be faster than the reduction of the radius of the blackhole. Under this assumption we have shown that the physical universe expands and beyond a critical time it will accelerate.
\par We are pursuing the goal to see whether we can numerically obtain the correct state of expansion without forcing any extra assumption about the nature of the blackhole. It would be also interesting to pursue a similar analysis with other types of blackholes.
\par  Acknowledgement: A. L. Choudhury would like to thank the Fulbright Committee for supporting his visit at the University of Dhaka where this research has been carried out. He wants to express his deepest appreciation for all the support he got from colleagues while  teaching in the Department of Physics at the University of Dhaka.}
\section{Appendix}
{\quad Since most of the time we would restrict ourselves with the calculation of the Ricci tensor, we would just show the essentials of these calculations in this appendix. We know that}
\begin{equation}
R_{tt}=R^t_{ttt}+R^{r_\alpha}_{t{r_\alpha}t}+R^{r_\beta}_{t{r_\beta}t}+R^{\theta_\alpha}_{t{\theta_\alpha}t}+R^{\theta_\beta}_{t{\theta_\beta}t}+R^{\phi_\alpha}_{t{\phi_\alpha}t}+R^{\phi_\beta}_{t{\phi_\beta}t}.
\end{equation}
{\quad From the definition of $R^{\alpha}_{\beta\gamma\delta}$, we find}
\begin{equation}
R^t_{ttt}=0.
\end{equation}
{Similarly}
\begin{equation}
R^{r_\alpha}_{t{r_\alpha}t}=-\frac {\ddot\alpha} {\alpha}+\frac {{\dot\alpha}{\dot\beta}{\beta}(r_\beta-2m)} {{r_\beta}f(r_\beta,\beta)},
\end{equation}
\begin{equation}
R^{r_\beta}_{t{r_\beta}t}=-\frac{\ddot\beta}{\beta}+\frac{{\dot\beta}^2 (r_\beta-2m)} {f(r_\beta,\beta)}-\frac{\beta^2 m^2 (r_\beta-2m)} {r_\beta^4 f(r_\beta,\beta}-\frac{m^2} {{r_\beta}^2 (r_\beta-2m)}-\frac{m(r_\beta-2m)} {r_\beta}^4,
\end{equation}
\begin{equation}
R^{\theta_\alpha}_{t{\theta_\alpha}t}=-\frac{\ddot\alpha} {\alpha}+ \frac{{\dot\alpha}{\dot\beta}\beta(r_\beta-2m)} {\alpha {r_\beta}f(r_\beta,\beta)},
\end{equation}
\begin{equation}
R^{\theta_\beta}_{t{\theta_\beta}t}=-\frac{\ddot\beta} {\beta}+\frac{{{\dot\beta}^2}(r_\beta-2m)} {f(r_\beta,\beta)}+\frac{m(r_\beta-2m)} {{r_\beta}^4},
\end{equation}
\begin{equation}
R^{\phi_\alpha}_{t{\phi_\alpha}t}=-\frac{\ddot\alpha} {\alpha},
\end{equation}
\begin{equation}
R^{\phi_\beta}_{t{\phi_\beta}t}=R^{\theta_\beta}_{t{\theta_\beta}t}.
\end{equation}
{As a result we get}
\begin{equation}
R_{tt}=-\frac{3\ddot\alpha} \alpha +\frac{2\dot\alpha\dot\beta(r_\beta-2m)}{\alpha{}f(r_\beta,\beta)}-\frac{3\ddot\beta} \beta+\frac{3{\dot\beta}^2(r_\beta-2m)} {f(r_\beta,\beta)}-\frac{\beta^2 m^2 (r_\beta-2m)} {r_\beta^4 f(r_\beta,\beta)}+A(r_\beta,m),
\end{equation}
{where}
\begin{equation}
A(r_\beta,m)=\frac{m(r_\beta-2m)} {r_\beta^4}-\frac{m^2}{r_\beta^2(r_\beta-2m)}.
\end{equation}
{In the above equations we have defined $f(r_\beta,\beta)$ by the Eq.(18). }

\section{Reference}
\begin{enumerate}
\item A. L Choudhury: Influence of collapsing matter on the envelopinge expanding universe; arXiv.gr-qc/0506009v1, 1 Jun 2005.
\item N. Bachcall, J. P. Ostriker, S. Perlmutter, and P. J. Steinhardt, Science, 284, 1481 (1999).
\item A. L. Choudhury and H. Pendharkar, Hadronic J. 24,275 (2001).
\item S. B. Giddings and A. Strominger, Nucl. Phys. B 307, 854 (1988).
\item D. H. Coule and K. Maeda, Class. Quant. Grav. 7,955 (1990).
\item A. L. Choudhury: Influence on physical universe by wormhole generated extra dimensional space; arXiv.gr-qc/0311043v1, 13 Nov 2003.
\item Je-An Gu and W-Y. P.Huang, arXiv.astro-ph/0112365v1, 31 Dec 2001. 
\end{enumerate}

\end{document}